\documentclass[aps,prl,twocolumn,superscriptaddress,showpacs]{revtex4}
\usepackage{}
\usepackage{mathrsfs}
\usepackage{amssymb}
\usepackage{txfonts}
\usepackage{tipa}

\usepackage{graphicx}
\usepackage{dcolumn}
\usepackage{bm}

\begin{document}


\title{Inelastic neutron scattering study of crystal field excitations of Nd$^{3+}$ in NdFeAsO}

\author{Y. Xiao}
\email[y.xiao@fz-juelich.de]{}
\affiliation{J\"{u}lich Centre for Neutron Science JCNS and Peter Gr\"{u}nberg Institut PGI, JARA-FIT, Forschungszentrum J\"{u}lich GmbH, D-52425 J\"{u}lich, Germany}

\author{M. Zbiri}
\affiliation{Institut Max von Laue-Paul Langevin, 6 rue Jules Horowitz, BP 156, 38042 Grenoble Cedex 9, France}

\author{R. A. Downie}
\affiliation{Institute of Chemical Sciences and Centre for Advanced Energy Storage and Recovery, School of Engineering and Physical Sciences, Heriot-Watt University, Edinburgh, EH14 4AS, United Kingdom}

\author{J. -W. G. Bos}
\affiliation{Institute of Chemical Sciences and Centre for Advanced Energy Storage and Recovery, School of Engineering and Physical Sciences, Heriot-Watt University, Edinburgh, EH14 4AS, United Kingdom}

\author{Th. Br\"{u}ckel}
\affiliation{J\"{u}lich Centre for Neutron Science JCNS and Peter Gr\"{u}nberg Institut PGI, JARA-FIT, Forschungszentrum J\"{u}lich GmbH, D-52425 J\"{u}lich, Germany}

\author{T. Chatterji}
\email[chatterji@ill.fr]{}
\affiliation{Institut Max von Laue-Paul Langevin, 6 rue Jules Horowitz, BP 156, 38042 Grenoble Cedex 9, France}

\date{\today}

\begin{abstract}

Inelastic neutron scattering experiments were performed to investigate the crystalline electric field (CEF) excitations of Nd$^{3+}$ (\emph{J} = 9/2) in the iron pnictide NdFeAsO. The crystal field level structures for both the high-temperature paramagnetic phase and the low-temperature antiferromagnetic phase of NdFeAsO are constructed. The variation of CEF excitations of Nd$^{3+}$ reflects not only the change of local symmetry but also the change of magnetic ordered state of the Fe sublattice. By analyzing the crystal field interaction with a crystal field Hamiltonian, the crystal field parameters are obtained. It was found that the sign of the fourth and sixth-order crystal field parameters change upon the magnetic phase transition at $\sim$140 K, which may be due to the variation of exchange interactions between the 4\emph{f} and conduction electrons.

\end{abstract}

\maketitle


The discovery of superconductivity at \emph{T}$_c$ = 26 K in LaFeAsO$_{1-x}$F$_x$ has triggered extensive research on the physical properties and the superconducting mechanism of Fe-based superconductors \cite{Kamihara,Johnston}. Superconducting transition temperatures higher than 50 K can be reached in \emph{R}FeAsO$_{1-x}$F$_x$ systems if one replaces La by rare-earth element \emph{R}, such as Pr, Nd and Sm \cite{Ren1,Chen1,Ren2, Kito}. The enhancement of \emph{T}$_c$ by replacing non-magnetic La with magnetic rare-earth ions is argued to be mostly due to the variation of geometric factors instead of the rare-earth magnetism \cite{Johnston,Lee,Maeter}. Nevertheless, the interplay between the Fe 3\emph{d} and rare-earth 4\emph{f} magnetism in Fe pnictides still plays an important role in determining the structural and physical properties of these materials. For instance, strong couplings between rare-earth and Fe magnetism are found in both CeFeAsO and SmFeAsO undoped parent compounds \cite{Maeter, Nandi}. As for NdFeAsO, the Fe spin-density-wave (SDW) transition at \emph{T}$_{SDW}$ = 137 K is found to be preceded by a tetragonal to orthorhombic structural phase transition at \emph{T}$_S$ = 142 K and a spin-reorientation of Fe magnetic sublattice takes place upon the ordering of Nd moments at low temperature \cite{Qiu,Tian,Chatterji,Marcinkova1}.

As a major interaction in rare earth compounds, CEF interaction reflects directly the electrical and magnetic potential on the rare-earth site which is created by neighboring ions \cite{Fulde,Jensen}. Experimental determination of CEF structure and CEF parameters is critical to get insight into the energy scale of CEF interaction and to enhance the understanding of the physical properties of rare-earth contained compounds based on the degeneracy of CEF states. As the CEF interaction in the NdFeAsO parent phase has to date not been investigated, we studied the crystal field excitation of Nd$^{3+}$ in NdFeAsO by inelastic neutron scattering. Based on the CEF analysis with single ion model, the crystal field levels are established and the CEF parameters are obtained for both the paramagnetic and antiferromagnetic phase of NdFeAsO. The ground state of the Nd$^{3+}$ ion was found to be a magnetic doublet in the high-temperature paramagnetic phase, while the molecular field generated by the long range order of Fe moments lifts the degeneracy further and results in a magnetic singlet ground state in the low-temperature antiferromagnetic phase. It was also found that both fourth- and sixth-rank crystal field interactions are significantly different due to the variation of rare-earth-conduction electron interactions in two different phases.


The inelastic neutron scattering measurements were performed on the direct-geometry thermal-neutron time-of-flight spectrometer IN4C at the Institut Laue Langevin (Grenoble, France). In order to cover an extended \emph{\textbf{Q}}-range and to gain in energy resolution, two incident wavelengths of $\lambda_i$ = 1.11 and 2.22 $\buildrel_\circ \over {\mathrm{A}}$ were selected using a pyrolitic graphite monochromator. About 7 g of NdFeAsO powder sample was prepared by standard solid state chemistry method as previously reported \cite{Marcinkova2}.  The sample was put inside a thin Aluminium sample holder that was fixed to the cold tip of the sample stick of a standard orange cryostat. Standard corrections including detector efficiency calibration and background substraction were performed. The data analysis was done using ILL software tools.


\begin{figure}[!h]
\includegraphics[width=8.5cm,height=14cm]{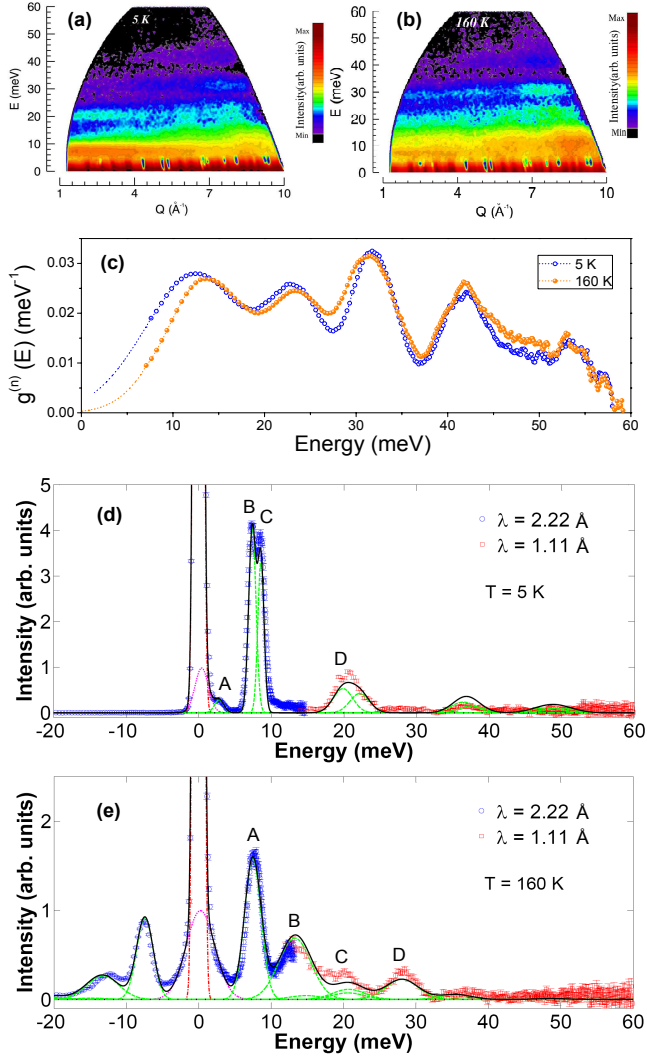}
\caption{\label{fig:epsart} (Color online) (a) and (b) The experimental Bose-factor corrected S(\emph{\textbf{Q}},$\omega$) plots for NdFeAsO at 5 and 160 K obtained using an incident neutron wavelength of 1.11 $\buildrel_\circ \over {\mathrm{A}}$. For clarity, a logarithmic representation is used for the intensities. (c) The experimentally generalized phonon density of states (GDOS) at 5 and 160 K for NdFeAsO. The GDOS was extracted within the incoherent approximation framework where the data was angle-averaged over the high-\emph{\textbf{Q}} region of the (\emph{\textbf{Q}},$\omega$) space of (a) and (b), i.e. 7 $\leq \textbf{Q} \leq$ 10 $\buildrel_\circ \over {\mathrm{A}} $$^{-1}$. (d) and (e) Energy spectra of CEF excitations for NdFeAsO at 5 and 160 K. The dash-dotted lines denote the elastic scattering line, the dotted line denotes the quasielastic scattering line shape fit, the dashed lines represent the fit of individual CEF transition as described in the text and the solid line is the fit of whole inelastic neutron scattering spectrum. The labeled peaks correspond to representative CEF transitions marked in Fig. 3(b).}
\end{figure}

Fig. 1(a) and (b) show the Bose-factor-corrected S(\emph{\textbf{Q}},$\omega$) plots for NdFeAsO compound at 5 and 160 K, respectively. Above \emph{T}$_{SDW}$, both the lattice and CEF excitations contribute to the spectra. Below \emph{T}$_{SDW}$, the neutron scattering intensity also contains the contribution from magnon scattering. Many inelastic neutron scattering studies on Fe pnictides have shown that the integrated cross sections of Fe spin wave excitations in pnictides are relatively small compared to that of phonon and CEF excitations \cite{Chi, Goremychkin}. Therefore, the CEF and lattice (phonon) excitations dominate the neutron spectra while the magnon scattering can be neglected during the data analysis, as presently done. The generalized phonon density of states (GDOS) was collected using an incident neutron wavelength $\lambda$$_i$ = 1.11 $\buildrel_\circ \over {\mathrm{A}}$. The GDOS (Fig. 1(c)) was extracted from the angle-integrated data over the high-\emph{\textbf{Q}} region of the (\emph{\textbf{Q}},$\omega$) space. The incoherent approximation is applied in the same way as in previous works dealing with phonon dynamics in Fe-pnictides \cite{Zbiri1,Zbiri2,Mittal1,Mittal2}. Our phonon spectra agree well with other measurements on NdFeAsO and NdFeAsO$_{1-x}$F$_x$ using inelastic x-ray scattering techniques \cite{LeTacon}. Further, the present GDOS shows similar features as those of the isostructural LaFeAsO$_{1-x}$F$_x$ compound obtained via inelastic neutron scattering measurements \cite{Christianson}. One can notice that upon the structural phase transition, the observed phonon features change only slightly. It is known that the phonon scattering intensity generally increases proportional to \emph{\textbf{Q}}$^2$. Pure CFE, which are not coupled to propagating modes, are local excitations and as such do not possess a characteristic dispersion and the scattering intensity decrease with \emph{\textbf{Q}} following a magnetic form factor. By taking advantage of the different \emph{\textbf{Q}}-dependence characters for phonon and CEF excitations, the contribution from CEF excitations can be separated by subtracting the phonon intensity from the experimental neutron spectra.

The obtained CEF excitation spectra at 5 K and 160 K are shown in Fig. 1 (d) and (e) as plots of the energy dependence of the scattering intensity. With increasing temperature, the intensities of all inelastic peaks in Fig. 1(d) are decreased, indicating that these peaks originate from the excitations between the ground state to excited states. As the temperature increases to 160 K, few new peaks emerged and their intensity increases with increasing temperature. These peaks can be identified as the CEF transitions between the populated excited states. Besides, the de-excitation transition from excited states to the ground state can also be observed on the energy gain side (negative energy) of the neutron spectrum (Fig.1(e)). The dynamic structure factors for neutron energy gain and neutron energy loss are related through the principle of detailed balance, i.e. $S(-\omega)=e^{-\hbar\omega/kT} S(\omega)$. To interpret the CEF spectra, we analyze the data in the framework of an ionic CEF model with Hamiltonian:
\begin{eqnarray}
\hat{H}  = \hat{H}_{CEF} +  2 \mu _B (g_J-1) B_{mol} \hat{J}
\end{eqnarray}

\noindent where $\hat{H}_{CEF}$ represents the crystalline electric field Hamiltonian, which describes the CEF interaction at the Nd$^{3+}$ site in NdFeAsO compound. The second term is the contribution of a molecular magnetic field coupling to the total angular momentum \emph{$\hat{J}$} of the Nd$^{3+}$ ion.

In the high-temperature tetragonal phase, the Nd$^{3+}$ cations are located at the 2\emph{c} site under the \emph{C$_{4v}$} local symmetry. The corresponding CEF Hamiltonian can be written as: $\hat{H}^{Tetr.}_{CEF} = B^{0}_{2} \hat{O}^{0}_{2} + B^{0}_{4} \hat{O}^{0}_{4} + B^{4}_{4} \hat{O}^{4}_{4} + B^{0}_{6} \hat{O}^{0}_{6} + B^{4}_{6} \hat{O}^{4}_{6}$, where the $B^{m}_{n}$ are the CEF parameters and the $ \hat{O}^{m}_{n}$ are the CEF Stevens equivalent operators as defined in Ref. \cite{Wybourne}. The crystal field of the \emph{C$_{4v}$} symmetry splits the 10-fold multiplets of a free Nd$^{3+}$ ion into three $\Gamma$$_6$ and two $\Gamma$$_7$ Kramers doublets. In the low-temperature orthorhombic phase, the Nd$^{3+}$ $^4$I$_{9/2}$ ground multiplets are splitted by the orthorhombic crystal field of the \emph{C$_{2v}$} symmetry into five Kramers doublets, namely five $\Gamma$$_5$ states. The CEF Hamiltonian reads as: $\hat{H}^{Orth.}_{CEF} = B^{0}_{2} \hat{O}^{0}_{2} + B^{2}_{2} \hat{O}^{2}_{2} + B^{0}_{4} \hat{O}^{0}_{4} + B^{2}_{4} \hat{O}^{2}_{4} + B^{4}_{4} \hat{O}^{4}_{4} + B^{0}_{6} \hat{O}^{0}_{6} + B^{2}_{6} \hat{O}^{2}_{6} + B^{4}_{6} \hat{O}^{4}_{6} + B^{6}_{6} \hat{O}^{6}_{6}$. In principle, four CEF excitation peaks are expected to be observed at 5 K. Any further splitting of CEF peaks in the orthorhombic phase would indicate that the molecular field, i.e. the effective exchange magnetic field $B_{mol}$ induced by the Fe magnetic sublattice, plays a vital role and the degeneracy is further lifted. The local structure surrounding the Nd$^{3+}$ ion at 5 K is illustrated in Fig. 2(a).

\begin{figure}
\includegraphics[width=8.5cm,height=7.5cm]{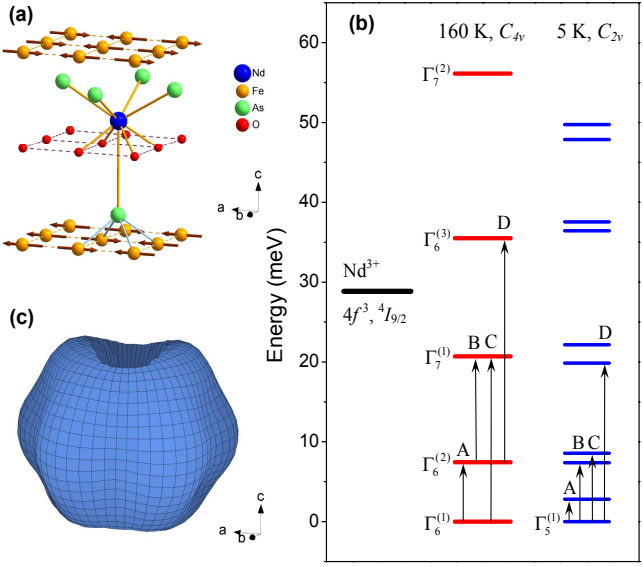}
\caption{\label{fig:epsart} (Color online) (a) Illustration of the local structure of the Nd$^{3+}$ ion at 5 K. It is subject to a crystal electric field generated by neighboring arsenic and oxygen ions, and a molecular field induced by the long range order of the iron spins. (b) Energy-levels scheme of Nd$^{3+}$ cations in NdFeAsO for the lowest \emph{J}-multiplets under the tetragonal and orthorhombic structure environments as derived from neutron spectroscopy. The values of energies at 160 K are sorted in increasing order as: 0, 7.45, 20.69, 35.51 and 56.12 meV. Similarly are shown the energies at 5 K: 0, 2.81, 7.38, 8.56, 19.86, 22.16, 36.44, 37.55, 47.86 and 49.75 meV. It is noted that excitations between any $\Gamma$$_6$ and $\Gamma$$_7$ states are allowed by the selection rules of neutron scattering at 160 K. The arrows denote a few representative CEF transitions as extracted from our neutron spectra. (c) The charge density distribution of Nd$^{3+}$ in NdFeAsO at 5 K.}
\end{figure}

Within the dipole approximation, the differential cross-section for unpolarized neutron scattering and for a CEF transition from state $| i \rangle$ to $| j \rangle$ can be expressed as:
\begin{eqnarray}
\frac{d^2 \sigma (i \rightarrow j)} {d\Omega dE'} =&&
N {\frac{k_f}{k_i}} \left( \frac{\hbar \gamma e^2}{mc^2} \right)^2 e^{-2W} \left\vert {\frac{1}{2}} g_J F(Q) \right\vert ^2  \nonumber\\
&&
\times  \sum_{i,j} n_i |\langle j| J_\bot | i \rangle ^2 \delta ( E_i - E_j +  \hbar \omega)
\end{eqnarray}
\noindent where \emph{k}$_i$ and \emph{k}$_f$ are the initial and final neutron wave vectors, and $\hbar \omega$ is the energy transfer. $\gamma$ is the neutron gyromagnetic ratio,  $e^2/mc^2$ is the classical electron radius, $g_J$ is the Land\'{e} factor, \emph{F(Q)} is the magnetic form factor, $e^{-2W}$ is the Debye-Waller factor, $n_i$ is the probability distribution of initial states, and $J_\bot$ is the component of the total angular-momentum operator perpendicular to \emph{\textbf{Q}}. By taking into account of the observed CEF energy positions, transition intensities and the evolution of the spectra as a function of temperature, a relevant fit is performed allowing to obtain reliable crystal field parameters that can be used to describe the behavior of the CEF excitations in NdFeAsO \cite{Program}. As indicated by the solid lines in Fig. 1(d) and (e), the results of the fitting lead to a good agreement with the observed spectra. Two sets of CEF parameters are obtained and listed in Table. 1, at 5 and 160 K, respectively. A schematic diagram of the splitting of the Nd$^{3+}$ $^4$I$_{9/2}$ ground multiplet is presented in Fig. 2(b). It is obvious that the ground state at 5 K is a magnetic singlet, whereas the ground state at 160 K is a magnetic doublet. Usually, the different ground states would result in different quasielastic scattering behavior. Compared with the neutron spectrum at 5 K (Fig. 1(d)), one prominent feature observed in the 160 K spectrum (Fig. 1(e)) is the appearance of considerable quasielastic peak intensity nearby the resolution limit elastic peak. The quasielastic signal can be understood as the consequence of spin fluctuations occurring within the degenerate ground state of Nd$^{3+}$ ions and it favors the multiplet ground state. The strong quasielastic peak observed at 160 K is mainly due to the magnetic doublet ground state with wave function $\psi_g$ = 0.2505$\mid$$\frac{9}{2}$, $\pm$$\frac{1}{2}$$\rangle$ + 0.2239$\mid$$\frac{9}{2}$, $\mp$$\frac{7}{2}$$\rangle$ + 0.9419 $\mid$$\frac{9}{2}$, $\pm$$\frac{9}{2}$$\rangle$, whereas the ground state at 5 K is a magnetic singlet with wave function $\psi_g$ = 0.9585$\mid$$\frac{9}{2}$, -$\frac{9}{2}$$\rangle$ -0.0403$\mid$$\frac{9}{2}$, -$\frac{7}{2}$$\rangle$ -0.1268$\mid$$\frac{9}{2}$, -$\frac{5}{2}$$\rangle$ +0.0144 $\mid$$\frac{9}{2}$, -$\frac{3}{2}$$\rangle$ +0.0255$\mid$$\frac{9}{2}$, -$\frac{1}{2}$$\rangle$  -0.0028 $\mid$$\frac{9}{2}$, $\frac{1}{2}$$\rangle$ -0.2467 $\mid$$\frac{9}{2}$, $\frac{3}{2}$$\rangle$ +0.0344  $\mid$$\frac{9}{2}$, $\frac{5}{2}$$\rangle$ +0.0198 $\mid$$\frac{9}{2}$, $\frac{7}{2}$$\rangle$ -0.0173$\mid$$\frac{9}{2}$, $\frac{9}{2}$$\rangle$.

\begin{table*}
\caption{\label{tab:table2} The obtained value of CEF parameters, in units of meV, for NdFeAsO at 5 and 160 K. The molecular fields obtained at 5 K are $B_{mol}^x$ = 11.2(3) T, $B_{mol}^z$ = 11.7(3) T.}
\begin{ruledtabular}
\begin{tabular}{cccccccccc}
&\emph{B}$^0_2$&\emph{B}$^2_2$&\emph{B}$^0_4$&\emph{B}$^2_4$&\emph{B}$^4_4$&\emph{B}$^0_6$&\emph{B}$^2_6$&\emph{B}$^4_6$&\emph{B}$^6_6$\\
\hline

5K &-0.567(9)&4.1(3)$\times$10$^{-2}$&1.1(2)$\times$10$^{-3}$&-8.8(3)$\times$10$^{-3}$&-3.2(2)$\times$10$^{-2}$&-8.6(5)$\times$10$^{-5}$&5.1(4)$\times$ 10$^{-4}$&6.7(4)$\times$10$^{-4}$&1.9(3)$\times$10$^{-3}$\\
160K&-0.584(9)& &-2.6(3)$\times$10$^{-3}$& &-5.7(3)$\times$10$^{-2}$&1.5(5)$\times$10$^{-4}$& & 4.6(5)$\times$10$^{-4}$&\\

\end{tabular}
\end{ruledtabular}
\end{table*}

Generally, the CEF parameters of insulating materials can be obtained from point charge model based calculations \cite{Hutchings}. It is assumed that the 4\emph{f} wave function do not overlap with those of neighboring ions, and the electrostatic crystal field potential is only created by the charge distribution of the neighboring ions. However, in metallic systems, the conduction electrons screen out the neighboring ionic point charge contribution, and the 4\emph{f}-conduction electron interaction plays an important role in the determination of CEF parameters. At the rare earth sites, the conduction and 4\emph{f} electrons interact via direct and exchange Coulomb interactions \cite{Levy, Schmitt1,Schmitt2}. It was found that the exchange coulombic contribution of conduction electrons to the CEF in rare-earth intermetallics may counterbalance the direct Coulomb contribution and change the sign and magnitude of the effective CEF parameters $B_4^0$ and $B_6^0$ \cite{Chow,Schmitt2}. By comparing the CEF parameters obtained for 5 and 160 K in NdFeAsO, it is noticed that the \emph{B}$^0_2$, \emph{B}$^4_4$ and \emph{B}$^4_6$ do not exhibit strong variation under the two different phases, whilst the \emph{B}$^0_4$ and \emph{B}$^0_6$ are subject to a change in both their sign and magnitude. It is suggested that the exchange contribution is larger in the antiferromagnetic phase than in the paramagnetic phase of the metallic NdFeAsO system due to the stronger overlap of the 4\emph{f} and conduction electron radial wavefunctions.

\begin{figure}
\includegraphics[width=8.5cm,height=7cm]{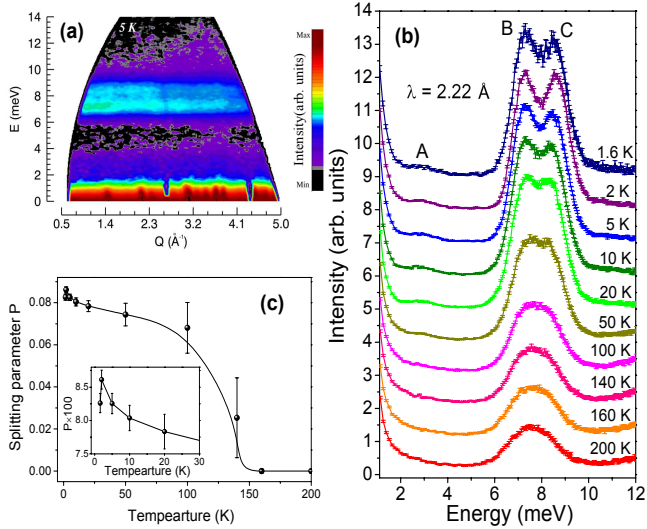}
\caption{\label{fig:epsart} (Color online) (a) The experimental Bose-factor corrected S(\emph{\textbf{Q}},$\omega$) plots for NdFeAsO at 5 K obtained using an incident neutron wavelength of 2.22 $\buildrel_\circ \over {\mathrm{A}}$. For clarity, a logarithmic representation is used for the intensities. (b) Temperature dependence of CEF excitations between 2 and 12 meV. (c) Temperature dependence of the peak splitting parameter. The lines are guides to the eye. Inset shows the enlarged view between 0 and 30 K.}
\end{figure}

By using the CEF parameters determined by inelastic neutron scattering measurements at 5 K, the spatial charge density distribution of the 4\emph{f} electrons of Nd$^{3+}$ in NdFeAsO was calculated \cite{Rotter}. As illustrated in Fig. 2(c), the charge density distribution is significantly nonspherical with slight variation in the \emph{ab} plane, and a large contraction along the \emph{c}-axis. The distortion of the charge density by CEF will lead to the magnetocrystalline anisotropy of the compound. Given the fact that Nd$^{3+}$ cations possess a negative second-order Stevens factor $\alpha_J$, the easy axis anisotropy of the Nd magnetic sublattice at 5 K is expected to be along the \emph{c}-axis. This is analogous to the observation in isostructural SmFeAsO and CeFeAsO compounds, where the moment directions of Sm and Ce atoms were determined to be along the \emph{c} axis between $T^{RE}_{N}$ and $T_{SDW}$ \cite{Maeter,Nandi}. The rare-earth magnetic order in this temperature range is indeed induced by the 3\emph{d}-4\emph{f} exchange interactions.

In order to get deeper insights into the effect of the molecular field of the Fe atoms on the CEF excitations, a S(Q,w) plot at 5K obtained from data with an incident wavelength of 2.22 $\buildrel_\circ \over {\mathrm{A}}$ giving a better resolution in the low energy region as shown in Fig. 3(a). Two excitation peaks at 7.24 and 8.59 meV are clearly observed and the intensity of these two peaks decrease following the square of the magnetic form factor \emph{F}(\emph{\textbf{Q}}). The evolution of the INS spectra as a function of temperature is shown in Fig. 3(b). As the observed CEF peak intensity is proportional to the ground state transition matrix elements, $|\langle \Gamma^{(j)}_5 | J_\bot | \Gamma^{(1)}_5 \rangle|^2$, the marked decrease in intensity of the peaks B and C hints at the decrease of the Van Vleck contributions to static magnetic susceptibility. Temperature dependence of the estimated splitting parameter, \emph{P} = [\emph{E}$_{(C)}$-\emph{E}$_{(B)}$]/[\emph{E}$_{(C)}$+\emph{E}$_{(B)}$], reveals that the SDW transition is at around 140 K, which is associated with a long-range antiferromagnetic order of the Fe moments. Additionally, as shown in the inset of Fig. 3(c), the splitting parameter reaches its maximum at 2 K and reduces below 2 K. The latter finding indicates that the local potential at the Nd$^{3+}$ sites is changed due to the change of the Fe molecular field upon the spin-reorientation transition of Fe in NdFeAsO \cite{Marcinkova1}. Given the fact that the Nd moments order antiferromagnetically below 2 K in polycrystalline NdFeAsO samples \cite{Qiu,Chatterji}, the effective exchange field due to the Nd magnetic sublattice will undoubtedly influence the crystal field levels of Nd$^{3+}$ below 2 K. A similar phenomenon has been reported for PrFeAsO \cite{Goremychkin}, in which a collapse of the splitting has been observed when the Pr sublattice orderes magnetically at \emph{T}$_N^{Pr}$ = 12 K.


In summary, we have performed temperature dependent inelastic neutron scattering measurements of the crystal field excitations in a NdFeAsO compound. With decreasing tempeature, the crystal field ground state of Nd$^{3+}$ changes from a magnetic doublet to a magnetic singlet state, accompanied with the occurrence of a long range magnetic ordering of the Fe moments. The crystal field parameters for the Nd$^{3+}$ ions have been obtained for both the high-temperature paramagnetic and the low-temperature antiferromagnetic phases, based on the analysis of inelastic neutron spectra with a single ion crystal field model. The significant difference in the high order CEF parameters between the two different phases reflects a great difference in the interaction between the rare-earth 4\emph{f} and conduction electrons.

\appendix


\begin{thebibliography}{10}

\bibitem{Kamihara}
Y. Kamihara, T. Watanabe, M. Hirano, and H. Hosono, J. Am. Chem.
Soc. \textbf{130}, 3296 (2008).

\bibitem{Johnston}
David C. Johnston, Advances in Physics \textbf{59}, 803(2010), and
references therein.




\bibitem{Ren1}
Z.-A. Ren, J. Yang, W. Lu, W. Yi, G.-C. Che, X.-L. Dong, L.-L. Sun, and Z.-X. Zhao, Mater. Res. Innovations 12, 105 (2008)..


\bibitem{Chen1}
X. H. Chen, T. Wu, G. Wu, R. H. Liu, H. Chen, and D. F. Fang, Nature (London) 453, 761 (2008).


\bibitem{Ren2}
Z.-A. Ren, J. Yang, W. Lu, W. Yi, X.-L. Shen, Z.-C. Li, G.-C. Che, X.-L. Dong, L.-L. Sun, and F. Zhou, Europhys. Lett. 82, 57002 (2008).

\bibitem{Kito}
H. Kito, H. Eisaki, and A. Iyo, J. Phys. Soc. Jpn. 77, 063707 (2008).

\bibitem{Lee}
C.-H. Lee, A. Iyo, H. Eisaki, H. Kito, M.T. Fernandez-Diaz, T. Ito, K. Kihou, H. Matsuhata, M. Braden, and K. Yamada, J. Phys. Soc. Jpn. 77 (2008), p. 083704.

\bibitem{Maeter}
H. Maeter, H. Luetkens, Y. G. Pashkevich, A. Kwadrin, R. Khasanov, A. Amato, A. A. Gusev, K. V. Lamonova, D. A. Chervinskii, R. Klingeler, C. Hess, G. Behr, B. B\"{u}chner, and H.-H. Klauss, Phys. Rev. B 80, 094524 (2009).

\bibitem{Nandi}
S. Nandi, Y. Su, Y. Xiao, S. Price, X. F. Wang, X. H. Chen, J. Herrero-Martin, C. Mazzoli, H. C. Walker, L. Paolasini et al., Phys. Rev. B 84, 054419 (2011).


\bibitem{Qiu}
Y. Qiu, Wei Bao, Q. Huang, T. Yildirim, J. M. Simmons, M. A. Green, J. W. Lynn, Y. C. Gasparovic, J. Li, T. Wu, G. Wu, and X. H. Chen,  Phys. Rev. Lett. \textbf{101}, 257002 (2008).

\bibitem{Tian}
W. Tian, W. Ratcliff II, M. G. Kim, J.-Q. Yan, P. A. Kienzle, Q. Huang, B. Jensen, K. W. Dennis, R. W. McCallum, T. A. Lograsso, R. J. McQueeney, A. I. Goldman, J. W. Lynn, and A. Kreyssig, Phys. Rev. B \textbf{82}, (R)060514 (2010).

\bibitem{Chatterji}
T. Chatterji, G. N. Iles, B. Frick, A. Marcinkova, and J.-W. G. Bos,  Phys. Rev. B 84, 132413 (2011).

\bibitem{Marcinkova1}
A. Marcinkova, T. C. Hansen and J. -W. G. Bos, J. Phys.: Condens. Matter \textbf{24},  256007 (2012).


\bibitem{Fulde}
P. Fulde and M. Loewenhaupt, Advances in Physics \textbf{34}, 589(1985).
\bibitem{Jensen}
J. Jensen and A. R. Mackintosh, \emph{Rare Earth Magnetism}, Clarendon Press Oxford (1991)

\bibitem{Marcinkova2}
A. Marcinkova,  E. Suard, A. N. Fitch, S. Margadonna and J. -W. G. Bos, Chem. Mater. \textbf{21}, 2967 (2009).



\bibitem{Wybourne}
B. G. Wybourne, \emph{Spectroscopic Properties of Rare Earths}, Interscience Pub., New York, (1965).


\bibitem{Chi}
S. Chi, D. T. Adroja, T. Guidi, R. Bewley, S. Li, Jun Zhao, J.W. Lynn, C. M. Brown, Y. Qiu, G. F. Chen, J. L. Lou, N. L. Wang, and P. Dai,  Phys. Rev. Lett. \textbf{101}, 217002 (2008).


\bibitem{Goremychkin}
E. A. Goremychkin, R. Osborn, C. H. Wang, M. D. Lumsden, M. A. McGuire, A. S. Sefat, B. C. Sales, D. Mandrus, H. M. R${\o}$nnow, Y. Su, and A. D. Christianson, Phys. Rev. B \textbf{83}, 212505 (2011).

\bibitem{Zbiri1}
M. Zbiri, R. Mittal, S. Rols, Y. Su, Y. Xiao, H. Schober, S. L. Chaplot, M. Johnson, T. Chatterji, Y. Inoue, S. Matsuishi, H. Hosono, and Th. Brueckel, J. Phys.: Condens. Matter \textbf{22} 315701 (2010).

\bibitem{Zbiri2}
M. Zbiri, H. Schober, M. R. Johnson, S. Rols, R. Mittal, Y. Su, M. Rotter, and D. Johrendt, Phys. Rev. B \textbf{79} 064511 (2009).

\bibitem{Mittal1}
R. Mittal, M. Zbiri, S. Rols, Y. Su, Y. Xiao, H. Schober, S. L. Chaplot, M. Johnson, T. Chatterji, S. Matsuishi, H. Hosono, and Th. Brueckel, Physical Review B \textbf{79} 214514 (2009).

\bibitem{Mittal2}
R. Mittal, S. Rols, M. Zbiri, Y. Su, H. Schober, S. L. Chaplot, M. Johnson, M. Tegel, T. Chatterji, S. Matsuishi, H. Hosono, D. Johrendt, and Th. Brueckel, Physical Review B \textbf{79} 144516 (2009).

\bibitem{LeTacon}
M. Le Tacon, M. Krisch, A. Bosak, J.-W. G. Bos, and S. Margadonna, Phys. Rev. B \textbf{78}, 140505(R) (2008).

\bibitem{Christianson}
A. D. Christianson, M. D. Lumsden, O. Delaire, M. B. Stone, D. L. Abernathy, M. A. McGuire, A. S. Sefat, R. Jin, B. C. Sales, D. Mandrus, E. D. Mun, P. C. Canfield, J.Y.Y. Lin, M. Lucas, M. Kresch, J. B. Keith, B. Fultz, E. A. Goremychkin, and R. J. McQueeney, Phys. Rev. Lett. \textbf{101}, 157004 (2008)

\bibitem{Program}
The CEF parameters are derived using SAFiCF program package and McPhase program package.

\bibitem{Hutchings}
M. T. Hutchings, ¡°Point Charge Calculations of Energy Levels of Magnetic Ions in Crystalline Electric Fields¡°, in: Solid State Physics Vol. 16, ed. F. Seitz and D. Thurnbull, Academic Press,New York and London (1964) 227




\bibitem{Levy}
P. M. Levy, Journal de Physique, \textbf{40}, C5-8 (1979).
\bibitem{Schmitt1}
D. Schmitt, J. Phys. F:Metal Phys., \textbf{9}, 1745(1979).

\bibitem{Schmitt2}
D. Schmitt, J. Phys. F:Metal Phys., \textbf{9}, 1759(1979).
\bibitem{Chow}
H. C. Chow, Phy. Rev. B, \textbf{7}, 3404(1973).


\bibitem{Rotter}
M. Rotter and A. T. Boothroyd, Phys. Rev. B \textbf{79}, 140405(R) 2009;
McPhase program, http://www.McPhase.de.





\end{thebibliography}
\end{document}